\documentclass[review]{elsarticle}
\graphicspath{ {./figures/} }
\usepackage{hyperref}
\usepackage{float}
\usepackage{verbatim} 
\usepackage{apalike}
\usepackage{xcolor}
\usepackage[numbers]{natbib}
\restylefloat{figure}
\restylefloat{table}
\usepackage{caption}
\usepackage{subcaption}
\usepackage{epsfig}

\usepackage{lscape}

\journal{}

\biboptions{authoryear}

\begin{document}
\begin{frontmatter}
\begin{titlepage}
\begin{center}
\vspace*{1cm}

\textbf{ \large Complex Network-Based Approach for Feature Extraction and Classification of Musical Genres}

\vspace{1.5cm}

Matheus Henrique Pimenta-Zanon$^{a,b}$ (matheus.pimenta@outlook.com), Glaucia Maria Bressan$^b$ (glauciabressan@utfpr.edu.br), Fabr\'icio Martins Lopes$^a$ (fabricio@utfpr.edu.br) \\

\hspace{10pt}

\begin{flushleft}
\small  
$^a$ Universidade Tecnol\'ogica Federal do Paran\'a (UTFPR), Computer Science Department, Alberto Carazzai, 1640, 86300-000, Corn\'elio Proc\'opio, PR, Brazil \\
$^b$ Universidade Tecnol\'ogica Federal do Paran\'a (UTFPR), Mathematics Department, Alberto Carazzai, 1640, 86300-000, Corn\'elio Proc\'opio, PR, Brazil 

\vspace{1cm}
\textbf{Corresponding Author:} \\
Fabr\'icio Martins Lopes\\
Universidade Tecnol\'ogica Federal do Paran\'a (UTFPR), Computer Science Department, Alberto Carazzai, 1640, 86300-000, Corn\'elio Proc\'opio, PR, Brazil \\
Tel: +55 (43) 3133-3872 \\
Email: fabricio@utfpr.edu.br

\end{flushleft}        
\end{center}
\end{titlepage}

\title{Complex Network-Based Approach for Feature Extraction and Classification of Musical Genres}

\author[label1,label2]{Matheus Henrique Pimenta-Zanon}
\ead{matheus.pimenta@outlook.com}

\author[label2]{Glaucia Maria Bressan}
\ead{glauciabressan@utfpr.edu.br}

\author[label1]{Fabr\'icio Martins Lopes \corref{cor1}}
\ead{fabricio@utfpr.edu.br}

\cortext[cor1]{Corresponding author.}
\address[label1]{Universidade Tecnol\'ogica Federal do Paran\'a (UTFPR), Computer Science Department, Alberto Carazzai, 1640, 86300-000, Corn\'elio Proc\'opio, PR, Brazil}
\address[label2]{Universidade Tecnol\'ogica Federal do Paran\'a (UTFPR), Mathematics Department, Alberto Carazzai, 1640, 86300-000, Corn\'elio Proc\'opio, PR, Brazil}

\begin{abstract}
Musical genre's classification has been a relevant research topic. The association between music and genres is fundamental for the media industry, which manages musical recommendation systems, and for music streaming services, which may appear classified by genres. In this context, this work presents a feature extraction method for the automatic classification of musical genres, based on complex networks and their topological measurements. The proposed method initially converts the musics into sequences of musical notes and then maps the sequences as complex networks. Topological measurements are extracted to characterize the network topology, which composes a feature vector that applies to the classification of musical genres. The method was evaluated in the classification of 10 musical genres by adopting the GTZAN dataset and 8 musical genres by adopting the FMA dataset. The results were compared with methods in the literature. The proposed method outperformed all compared methods by presenting high accuracy and low standard deviation, showing its suitability for the musical genre’s classification, which contributes to the media industry in the automatic classification with assertiveness and robustness. The proposed method is implemented in an open source in the Python language and freely available at \url{https://github.com/omatheuspimenta/examinner}.
\end{abstract}

\begin{keyword}
Musical genres classification \sep expert music systems \sep feature extraction \sep pattern recognition \sep machine learning
\end{keyword}
\end{frontmatter}

\section{Introduction}
\label{sec:introduction}

Musical genres classification has become a relevant topic, especially for the media industry. The correct association between genres and musics is very important for several applications, for example, musical recommendation systems. The Music Information Retrieval (MIR) is a research topic regarding the music audience and consumption mainly from online sources containing millions of songs \citep{oramas2016exploring}.
Providing properly selected music contents is fundamental for the media industry, which manages huge music catalogues stored on distributed databases \citep{Fernandez}. The automatic classification of musical genres is one of the most popular tasks in MIR research \citep{Tzanetakis}, which addresses the development of techniques to analyze, organize, synthesize or extract musical information \citep{Ras}. 

The classification in genres aims to group music styles according to common properties and features. This classification task has motivated studies specially because of the improvement of digital technologies and computational power available. Therefore, automatic classification of music genres plays an essential role in music indexing and retrieval, allowing Web sites and expert music indexing systems to manage and label music content \citep{correa2016}. As more music is made available, the need for efficient methods to query and retrieve information from these musical databases increases \citep{Dannenberg}. 

According to current literature, research works dedicated to the music recovery considering its genre are based on learning strategies involving the feature extraction, which composes a feature vector \citep{duda2001}. The feature extraction is a decisive step for the success of the classification, since it deals with the reduction of data dimensionality, i.e. it aims to represent the original data in a smaller vector trying to avoid the loss of information. Therefore, the feature vectors are applied in classification methods and, as a result, an automatic classification system for musical genres. 

Although many techniques have already been proposed for music genre classification \citep{Sturm3,scaringella2006automatic,fu2011}, no general solution is available, mainly because of the imprecise definition of musical genres and merges among borders \citep{BressanSilla,fernandez2012}.
In this scenario, this work presents a feature extraction method based on complex networks for the automatic classification of musical genres, called EXAMINNER (EXtrAction of MusIcal Notes from complex NEtwoRks). Competitor methods are adopted in order to compare the proposed approach by considering accuracy, robustness, and methodological complexity. For that, the GTZAN \citep{Tzanetakis} and FMA \citep{fma} datasets were adopted, since they are commonly used in the literature and then the performances could be compared with other feature extraction methods.

The novelty of this work is a feature extraction method based on complex networks for musical genres identification by considering the sequence of musical notes from each considered music. The proposed method differs from those found in the literature, which use Wavelet, Fourier, Cossine, spectral and other transform, which increase the methodological complexity of the feature extraction \citep{Nanni2017, Panagakis, Silla}. In addition, some methods produce results in an embedded form that do not allow for an interpretation of how their results were generated (interpretability), such as the deep learning based methods \citep{Sigtia, Choi,Bisharad}. The proposed method considers only the sequences of musical notes, an information that can be retrieved directly from the music itself, with no additional information. Furthermore, the EXAMINNER is a feature extraction method that can be used independently of the classification algorithm, i.e. it is not embedded with the classification method, leading to a more general and non classifier-dependent solution, unlike other methodologies that are embedded with the classification algorithm \citep{turnbull2005fast,Iloga2014,oramas2018multimodal,Bisharad}.

The contributions that can be highlighted in this work are: new feature extraction method based on complex networks measurements, the high accuracy presented by the classification results, interpretability of its results, relative simplification (using only two parameters) with lower methodological complexity and higher robustness, when compared with competitor methods. In addition, the proposed approach can help the media industry in the musical genres classification task, offering automation services with robustness and confidence in its results.

The paper is organized as follows. Section 2 described the related recent approaches proposed in the literature. Section 3 presents the material and methods, detailing the proposed approach and the adopted datasets. The results and discussion are presented in Section 4, as well as the evaluation of the proposed approach and literature methods. Finally, the conclusion is presented in Section 5.

\section{Related Works}

Considering the context of music genre classification, some approaches have been proposed in the literature. In particular, the survey by \cite{Sturm3} presents the musical genre recognition and analyzes three major aspects: experimental designs, datasets and figures of merit. An overview of most important approaches that deal with music genre classification is also presented in the survey by \cite{correa2016}, which considers the symbolic representation of music data, presenting the current issues inherent to the music format and the main algorithms used to model the music feature space.

A multi-task transfer framework for using artist labels to improve a genre classification model is presented in \cite{kim2018}. Artist labels are adopted as side information, allowing a model to learn the mapping between audio and artists, while capturing patterns that might as well be useful for genre prediction. Authors show that music representations learned from raw artist labels can transfer to other music-related tasks.

In \cite{rosner2018}, authors investigated if separating music tracks and extending feature vector by parameters related to the specific musical instruments that are characteristic for the given musical genre allows for efficient automatic musical genre classification. Authors have showed that the feature vector and the SVM with Co-training mechanism are applicable to a large dataset, which was a subset of audio excerpts extracted from the Synat dataset.

According to \cite{Bisharad}, some features provide a satisfactory level of performance only on a particular dataset, however not on other datasets. In their study, authors proposed a music genre recognition using a convolutional recurrent neural network architecture (CRNN), which is trained on mel spectrogram, which is a simpler, lower level acoustic representation of audio signals \citep{shen2018natural}. Then, it is extracted audio clips of 3 seconds from the mel spectrogram and a convolutional network is trained to perform artist recognition, genre recognition and key detection \citep{Dieleman} by training a convolutional deep belief network on all data, and then use the learned parameters to initialize a convolutional multilayer perceptron with the same architecture. In addition, \cite{Choi} concatenated feature vectors using the activations of feature maps of multiple layers in a trained convolutional network. 
On the other hand, \cite{Sigtia} examines ways to improve feature learning for audio data using stochastic gradient descent and deep neural networks. The methods provide improvements in training time and music features.

Music genre classification is also addressed in a multilinear perspective \citep{Panagakis}, in which a multiscale spectro-temporal modulation features are extracted. Then, the classification is performed by a Support Vector Machine and a stratified cross-validation method. More recently, \cite{Panagakis2} proposed a music genre classification framework that combines the rich, psycho-physiologically grounded properties of auditory cortical representations of music recordings and the power of sparse representation-based classifiers. \cite{Panagakis3} proposed a music genre classification framework that combines the rich, psycho-physiologically grounded properties of slow temporal modulations of music recordings and the power of sparse representation-based classifiers. It is worth mentioning that \citep{Sturm2} presents a rigorous analysis of these systems, questioning their performance.

For the classification task, \cite{BressanSilla} present the construction of a fuzzy system for the automatic classification of Latin music genres, considering the imprecise definition and merges among borders of musical genres. For this, it adopted the selected features by \cite{Silla}.

The Markov models (MM) were adopted as classifiers to perform music genres classification by \cite{Iloga2014}. The aim of this approach is to capture the inner transitions from one genre to another in order to perform the classification. More specifically, the use of one MM per genre model statistically captures the transitions between genres through its state's transition matrix.

The approach for music genre classification presented by \cite{nanni2016} is based on the fusion of acoustic and visual features, considering an ensemble of SVM classifiers. Also, combining visual and acoustic features for automated audio classification, \cite{Nanni2017} use the input signal represented by its spectrogram and an ensemble of SVM classifiers. Existing music transcription methods normally perform pitch estimation. A method based on deep neural networks for audio-to-score music transcription of monophonic excerpts is proposed by \cite{roman2020}, which outputs a notation-level music score, using an audio file as input, modeled as a sequence of frames.

In the literature, several works that address musical genres classification adopt the librosa library \citep{paperlibrosa2015} as the feature extraction method. The librosa is a python package for audio and music signal processing, currently in version 0.8 \citep{pacote0.8librosa}. The following features have been extracted using the librosa library and adopted in the literature: Chroma (\cite{Goto2006, Muller2011}), Mel-frequency cepstral coefficients (MFCCs) (\cite{rabiner1993fundamentals}), root-mean-square (RMS) value for each music frame, spectral measures (\cite{klapuri2007signal, Dubnov2004}), coefficients to a nth-order polynomial to the columns of a spectrogram, tonal centroid features (\cite{Harte2006}), tempogram (\cite{Grosche2010}) and zero-crossing rate of a music time series.
These features are adopted in musical genres classification in \cite{fma} in order to generate a lower-bound to this task and show the task's difficulty to classify the musical genres in the FMA dataset.

In \cite{ForoughmandArabi2009}, authors adopt MFCC, Spectral Centroid, SpectralRolloff, Spectral Flux, and Zero Crossing features in order to classify musical genres, showing that the combination of low-level features with high-level features is effective. Therefore, the librosa library is applied by methods of identifying musical genres.

This work presents a complex-network based approach for feature extraction and music genre classification. Complex networks have been successfully applied to analyze, represent and understand complex systems in many application areas \citep{watts1998collective,newman2003,vazquez2004topological,BOCCALETTI,costa2007a,backes2009complex,backes2010complex,barabasi2011,Vicente2014,lopes2014a,Lima2015,Basinet,lima2019,Montanini2020}, contributing as a multidisciplinary methodology.


\section {Material and Methods}
\label{sec:mm}

\subsection{Material}
\label{subsec:material}

The dataset presented by \cite{Tzanetakis}, known as GTZAN dataset, contains $1,000$ music clips distributed across 10 classes of musical genres. A rigorous analysis about GTZAN dataset is presented in \cite{Sturm2012}, which shows that GTZAN has rhythms overlapping and repetition of musics. More specifically, it is reported that about 7\% of the overlapping of GTZAN come from the same recording (including 5\% being exact duplicates). Despite these issues inherent to the database, it is an important contribution for the literature, since many works adopt it in the assessments and the performance comparisons. Thus, this challenge is assumed in this work, which makes the assessment of the proposed method more realistic to this type of classification problem.

Thus, in order to evaluate the EXAMINNER feature extraction approach, the GTZAN dataset \citep{Tzanetakis} was adopted in this work. More specifically, the $1,000$ music clips are classified into ten music genres, namely: blues, classical, country, disco, hip hop, jazz, metal, pop, reggae, and rock. Each musical genre contains 100 samples with approximately $30$ seconds in duration of each one.

The second dataset adopted to evaluate the proposed method is the Free Music Archive (FMA) dataset \citep{fma}. The FMA dataset is mainly used to classify musical genre and data analysis. Thus, in order to evaluate the EXAMINNER feature extraction approach, the FMA small subset \citep{fma}, which is used for single label classification, was adopted in this work. This subset contains $8,000$ music files sorted in 8 musical genres, namely: electronic, experimental, folk, hip hop, instrumental, international, pop and rock. For each musical genre, $1,000$ balanced music files with 30 seconds in duration each one composes this dataset.

According to \cite{Gjerdingen}, humans can identify a musical genre with high accuracy only by listening 3 seconds of each music. This approach was successfully adopted in CRNN \citep{Bisharad}. Thus, in this work was adopted $10$ segments of $3$ seconds for each music.

As the result of the pre-processing stage, approximately $10$ segments with $3$ seconds in duration are got for each music. For each one of these segments, the fundamental frequency and musical notes are extracted. All the computational implementations are developed using the Python language. For the pre-processing stage, the {\it librosa} $0.8$ \citep{pacote0.8librosa} library was adopted.

\subsection{Methods}
\label{subsec:methods}

The EXAMINNER method is inspired on BASiNET \citep{Basinet}, which deals with the feature extraction for the classification of distinct classes of RNA sequences in bioinformatics research. It is discussed that the structure of nucleotides organization in the RNA sequence is an important factor in defining the classes of the sequences. More specifically, the neighborhood organization between the words of size 3 (k-mers, k=3) recovers the structural organization in the biological sequence, which is mapped in a complex network and its topological measurements are extracted to compose a feature vector, one for each sequence and, as a result, its classification with significant results.

\begin{figure*}[!ht]
	\centering
    \includegraphics[width=0.78\linewidth]{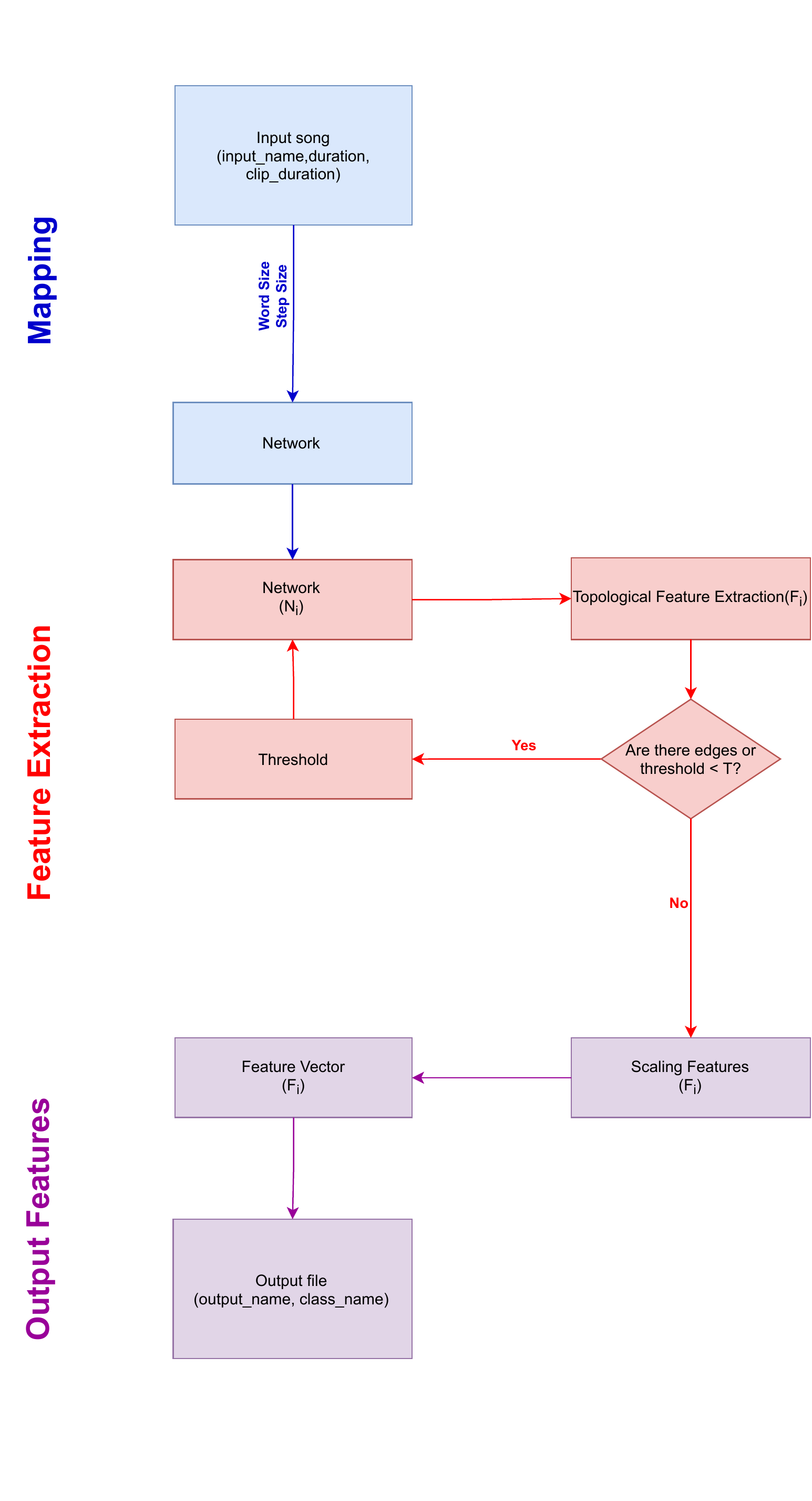}
	\caption{Overview of the EXAMINNER feature extraction method.}
	\label{fig:overview}
\end{figure*}

Therefore, considering that musics can be represented by sequences of musical notes and that the structure formed by the order in which the musical notes occur are relevant factors for its classification as a music genre, the EXAMINNER feature extraction is proposed, considering three main steps: mapping, feature extraction and output features. Figure \ref{fig:overview} presents an overview of the EXAMINNER feature extraction approach.

The EXAMINNER starts by considering each music individually. For each music, the fundamental frequency is extracted adopting the method proposed by \cite{Cheveigne} with the parameters of minimum frequency and maximum frequency equal to $65$Hz and $2093$Hz respectively, as recommend \citep{paperlibrosa2015,pacote0.8librosa}. After obtaining the fundamental frequency, the sequence of musical notes is obtained. Then, the musical notes of the chromatic scale are considered. For those notes the symbol \# was not considered, generating musical notes represented by two symbols.

\begin{figure*}[ht]
	\centering
	\includegraphics[width=0.9\linewidth]{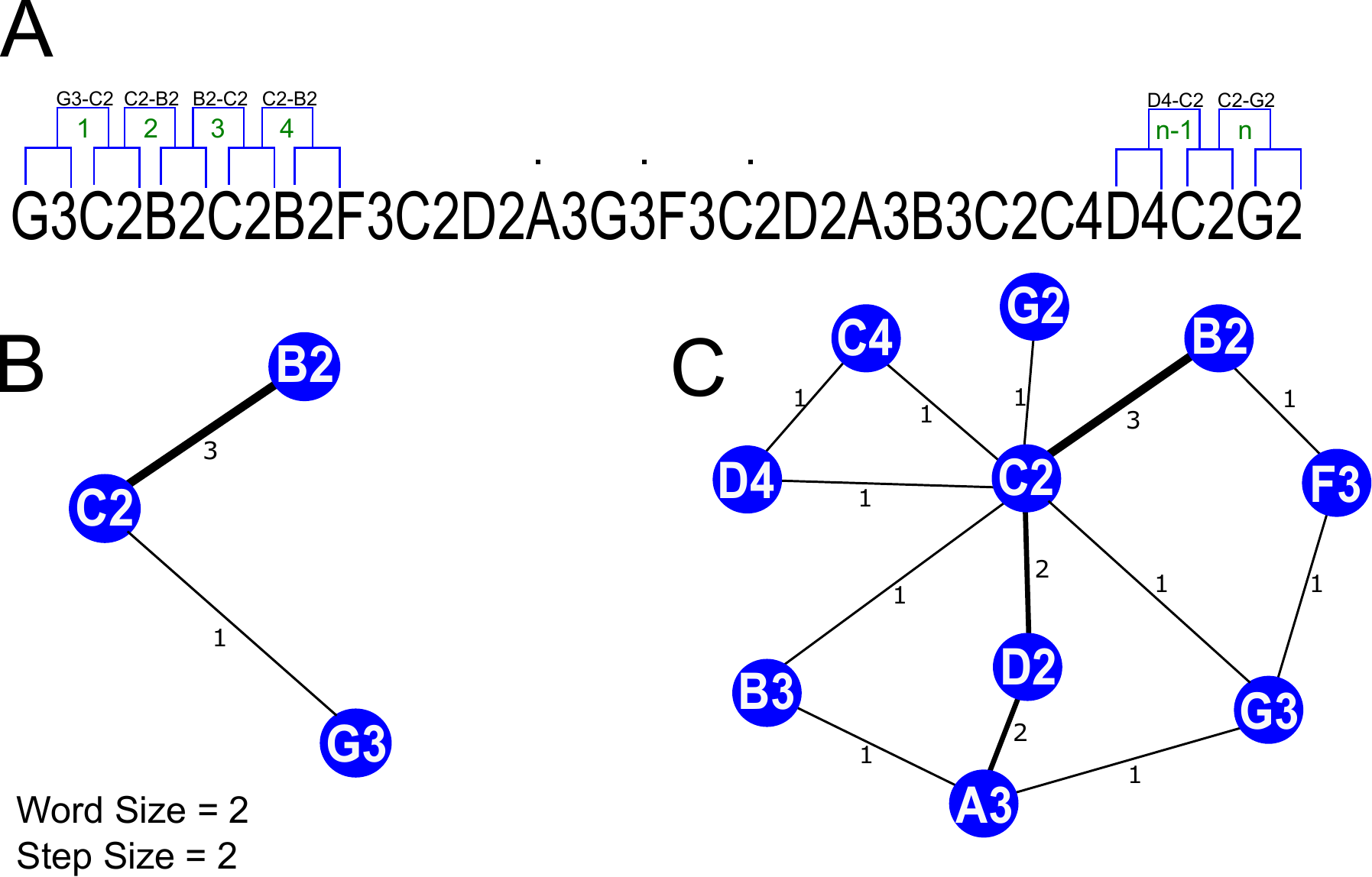}
	\caption{Mapping the sequence of musical notes to an undirected weighted network. In sub-figure {\bf (A)} the four first iterations of the algorithm with $WS=2$ and $ST=2$. The first iteration the node $G3$ is connected to node $C2$. In the next iteration with $ST=2$, the node $C2$ is connected to node $B2$ and the window slides with $ST=2$ size. Iteration $3$: the node $B2$ is connected to node $C2$ and the window slides with $ST=2$. The sub-figure {\bf (B)} network built from the first four illustrative iterations in sub-figure {\bf (A)}. The sub-figure {\bf (C)} shows the result network after all $n$ iterations.} 
	\label{fig:mapping}
\end{figure*}

Then, each sequence of musical notes is traversed and mapped into a complex network, with the objective of mapping the structural neighborhood relations of the musical notes in the music. Two parameters are considered for traversing the sequence: the word size ($WS$) and the step size ($ST$). In order to maintain coherence, this work adopts $WS=2$ and $ST=2$. More specifically, these values represent that the sequence of musical notes is traversed considering the immediate neighborhood and each note in the music is considered once. Figure \ref{fig:mapping} illustrates how these parameters are applied and the network is generated.

Therefore, the network is composed of musical notes as nodes, and the neighborhood relations between notes as edges. As a result, a complex network with undirected weighted edges is generated, in which its edges have an associated weight (numerical value) that represents the frequency of repetitions between the musical notes adjacency.

\begin{figure*}[!ht]
	\centering
	\includegraphics[width=0.8\linewidth]{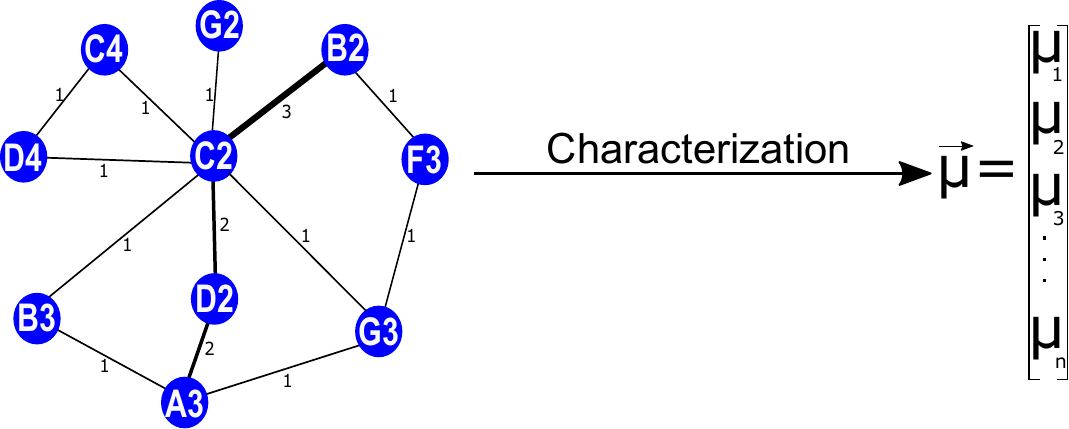}
	\caption{The mapping from a complex network into a feature vector.}
	\label{fig:features_vector}
\end{figure*}

The second main step is the feature extraction, in which complex network measurements \citep{costa2007a,BOCCALETTI} are extracted in order to characterize the organizational structure of musical notes in each music, and, as a result generating a feature vector as shown in Figure \ref{fig:features_vector}. More specifically, the initial network was considered with all the edges identified in mapping step and it is performed the topological feature extraction. Thus, 10 topological measurements commonly used in the literature were adopted in order to characterize the topological patterns of the network: assortativity (ASS), average degree (DEG), maximum degree (MAX), minimum degree (MIN), average betweenness centrality (BET), clustering coefficient (CC), average short path length (ASPL), average standard deviation (SD), frequency of motifs with size 3 (MT3) and frequency of motifs with size 4 (MT4).

However, the initial network has edges with different weights, i.e. the frequency of neighborhood between musical notes. Then the proposed approach iteratively applies a threshold to remove the ``weaker'' edges at each iteration. Removing edges will change the topology of the network and initially remove possible noise (less frequent edges) while keeping the most frequent (representative) edges, leading to a dynamic in the topological variation of the network. Topological measurements are extracted at each iteration, i.e. different frequency scales of the network edges. Therefore, the goal is to extract topological measurements at these different scales of connectivity between the nodes to compose a feature vector for each sequence of musical notes.
Then, thresholds are applied as long as there are edges left in the network (default) or until a $T$ value is reached, as an optional stopping condition. The number of thresholds $T$ is an optional parameter for the proposed method.

The third main step is the output of the feature vector. In order to avoid an inappropriate influence of the extracted features on the classification methods, i.e. different numerical scales of each feature, a Min-Max rescale is adopted. Thus, each feature vector $\vec{\mu} = {\mu_1, \mu_2, \ldots, \mu_n}$, the Min-Max re-scaling a value $\mu_k$ to $\mu_{rk}$ in the range $[0,1]$  defined as $\mu_{rk} = (\mu_k - \mu_{kmin}) / (\mu_{kmax} - \mu_{kmin})$. Where $\mu_{kmax}$ is the maximum value and $\mu_{kmin}$ is the minimal value for each $k$ topological measurement adopted. As a result, all the topological measurements are defined into the interval $[0,1]$.

Therefore, the normalized feature vector represents the output of the EXAMINNER feature extraction method, which can be applied by different classification algorithms \citep{duda2001}.

\section{Results and Discussion}
\label{sec:results}

Regarding the assessment of the EXAMINNER feature extraction method, the GTZAN and the FMA datasets were adopted. As described in Sec.~\ref{subsec:material}, 10 segments of 3 seconds were extracted for each music from {GTZAN} and FMA datasets in order to evaluate the proposed method and to compare with results from literature. Thus, considering the same input datasets, both EXAMINNER (Sec.~\ref{subsec:methods}) and librosa \citep{paperlibrosa2015} were applied for feature extraction. Regarding librosa, all features available by the library were considered and extracted.

In this context, the results section was organized in three parts: the results comparison of the proposed method with the librosa; comparison of the results of music genre classification using EXAMINNER compared to methods in the literature; and the analysis and interpretation of the features extracted from EXAMINNER and their use in music genre classification.

\subsection{Comparing EXAMINNER with librosa feature extractor}

In order to evaluate the proposed feature extraction approach, the librosa library \citep{paperlibrosa2015} was adopted to extract the musical features from the GTZAN and FMA datasets.
For each music, all the available features have been extracted from the librosa library feature extraction, as cited in Sec. 2: Chroma, Mel-frequency cepstral coefficients (MFCCs), root-mean-square (RMS) value for each music frame, spectral measures, coefficients to a nth-order polynomial to the columns of a spectrogram,  tonal centroid features, tempogram and zero-crossing rate of a music time series.

The EXAMINNER and librosa features were extracted by considering the same files, i.e. 3 sec segments. Regarding the librosa, for each feature it was also considered the mean and the standard deviation by using the numpy python library \citep{harris2020array}. This procedure was adopted for GTZAN and for FMA datasets.

In order to assess and to analyze the behavior of the EXAMINNER feature extraction method, several values of thresholds $T$ were adopted in the Feature Extraction step (see Fig. \ref{fig:overview}). In the first iteration, no threshold ($T0$) is considered, which produces a feature matrix with 10 columns (features) from the initial network, i.e. considering all the edges of the mapping step. In the next iteration, one threshold ($T1$) is considered and 10 more features are obtained, resulting in 20 features aggregated in the feature matrix, i.e. features $T0+T1$. This procedure was repeated iteratively, adding one by one threshold, until the tenth threshold ($T10$). After that, the next adopted thresholds were $T15$, $T20$, $T25$ and the last considered threshold was $T30$.

In addition, different classification algorithms were adopted in order to assess the extracted features, such as Classification via Regression (Regression), Bayes Network (BN), Random Forest (RF), support vector machine (SVM) and Multi Layer Perceptron (MLP) from Weka data mining software \citep{hall2009weka,lang2019wekadeeplearning4j} by adopting their default parameters, providing equality in the comparison of results. The 10-fold cross validation method was adopted \citep{classification}.

Tables \ref{tab:sup1} and \ref{tab:sup2} show the results by considering different threshold values for EXAMINNER feature extraction method. Table \ref{tab:sup1} shows the average accuracy of cross validation for each classifier and adopted threshold by considering GTZAN dataset and Table \ref{tab:sup2} shows the same considering FMA dataset. It is possible to notice that EXAMINNER produce higher accuracy rates, even adopting just 10 features ($T0$) from both GTZAN and FMA datasets. As new thresholds are applied and more features are included in the feature matrix, all the results of the classifier algorithms increase their accuracy for GTZAN dataset. A similar behavior occurs for FMA dataset, however with Random Forest, MLP and Regression classifiers showing a saturation and a slight loss of accuracy when considering the increasing number of features. As expected, the SVM classifier performs better at higher dimensional feature space \citep{cortes1995support} for both datasets.

\begin{figure*}[!ht]
    \centering
	\begin{subfigure}[b]{\textwidth}
	\centering
	\includegraphics[scale=0.7]{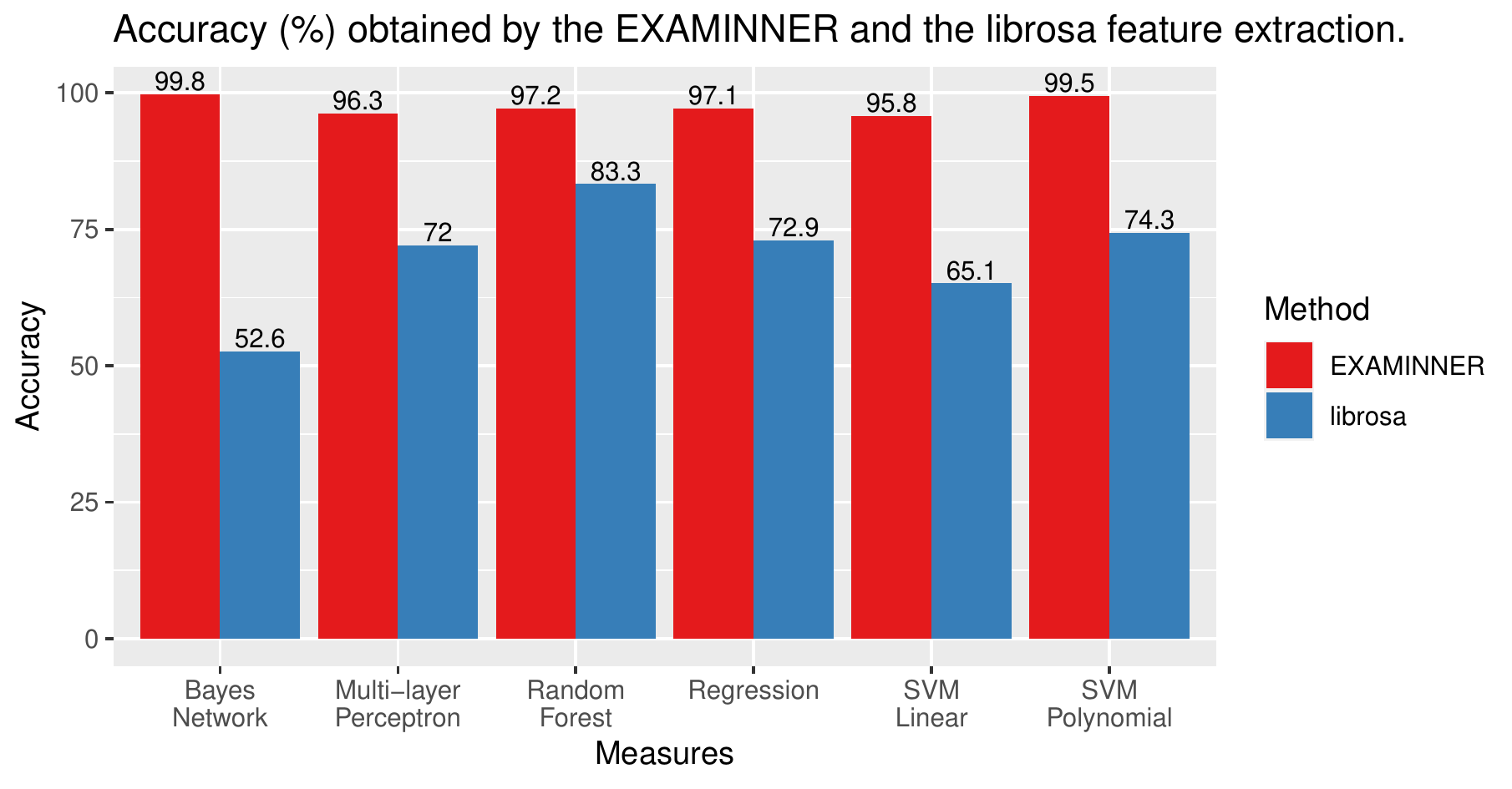}
	\caption{GTZAN dataset}
	\end{subfigure}
	\newline
	\begin{subfigure}[b]{\textwidth}
	\centering
	\includegraphics[scale=0.7]{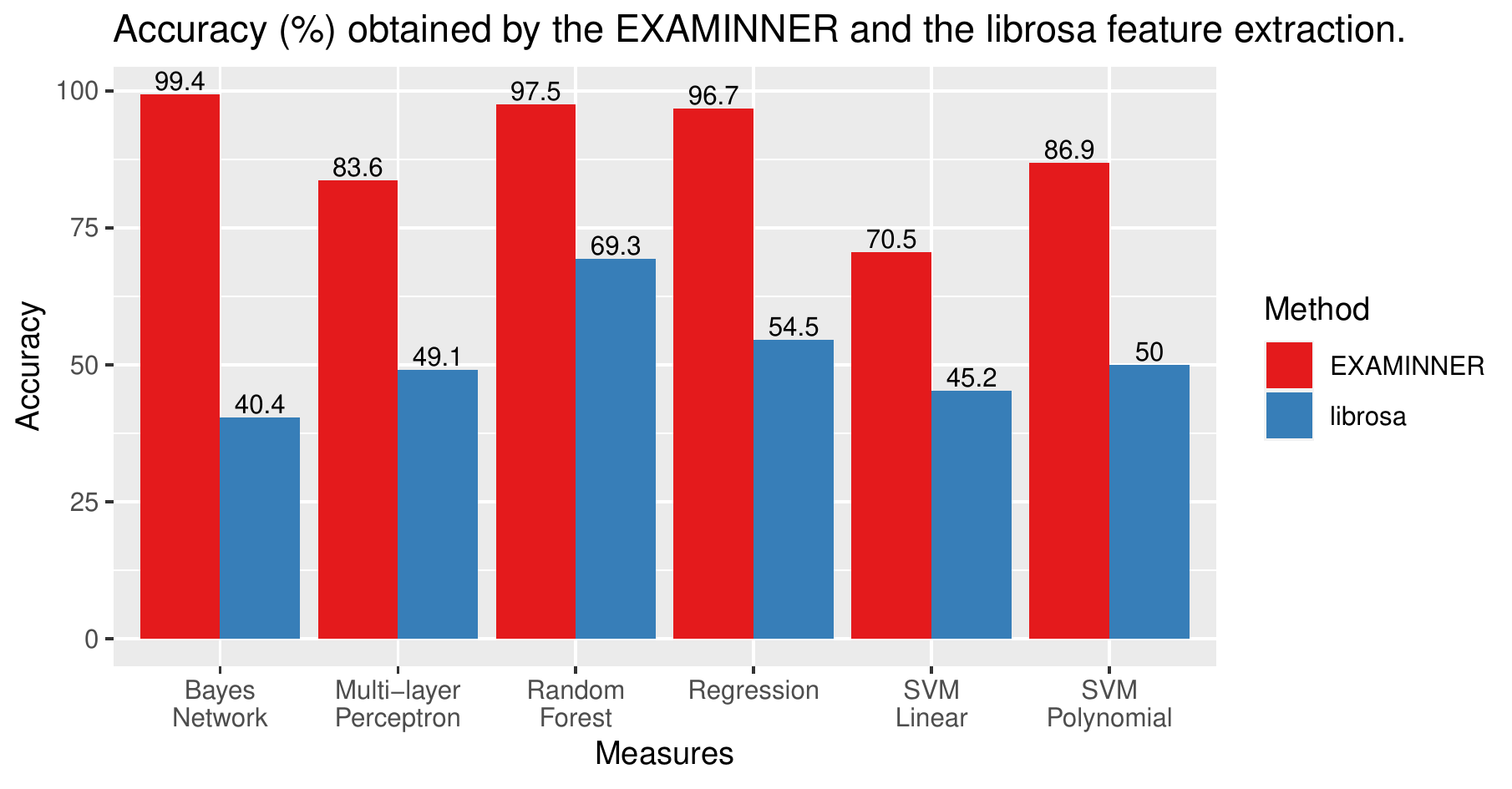}
	\caption{FMA dataset}
	\end{subfigure}
	\caption{The highest average accuracy of the 10-fold cross validation achieved from the adopted thresholds and classifiers when considering the features extracted by EXAMINNER compared to the features extracted by librosa library.}
	\label{fig:gtzanresult}
\end{figure*}

Figure \ref{fig:gtzanresult} shows the highest average accuracy values obtained for each classifier from EXAMINNER features comparing with librosa features, by considering different classifiers, respectively to (a) GTZAN dataset and to (b) FMA dataset.

Regarding the GTZAN dataset (Figure \ref{fig:gtzanresult}(a)), it is possible to observe that the behavior of all classifiers using EXAMINNER features was relatively homogeneous with lower variation, achieving higher accuracies for all classifiers algorithms than those achieved with the librosa features. The higher average accuracy was obtained by the Bayes Network classifier with 30 features from the EXAMINNER (threshold $T2$). Regarding the FMA dataset (Figure \ref{fig:gtzanresult}(b)), it is possible to notice that 
Bayes Net, Random Forest and Regression classifiers had better performance when compared with Multi-Layer Perception (MLP), Random Forest (RF) and Support Vector Machine (SVM) classifiers using the EXAMINNER features and the higher average accuracy was achieved by the Bayes Network classifier with 20 features from the EXAMINNER (threshold $T1$). It can be noted that all the classifiers produced higher average accuracy from the EXAMINNER features than the librosa, which considers 28 features.
Considering the results achieved by classifiers from librosa features, all of them present relatively lower accuracies when compared to the EXAMINNER, even considering the initial thresholds with less quantity of features than the librosa.

Although the tuning of the classification algorithms is not the goal of this work, SVM classifier presents the highest accuracy variation, thus the extracted features from librosa and EXAMINNER was also assessed by the SVM with a polynomial kernel (second degree). This evaluation shows, as expected, that the customization of the classifier improves the results already obtained.

In order to better understand the behavior of the classification algorithms, the confusion matrix is an important indicator for evaluating the results in each music genre, since it contains the number of elements that have been correctly or incorrectly classified for each class. The main diagonal presents the number of samples that have been correctly classified for each class. The off-diagonal elements present the number of samples that have been incorrectly classified \citep{duda2001}. 

\begin{figure*}
    \centering
    \begin{subfigure}[b]{0.475\textwidth}
    \centering
    \includegraphics[scale=0.3]{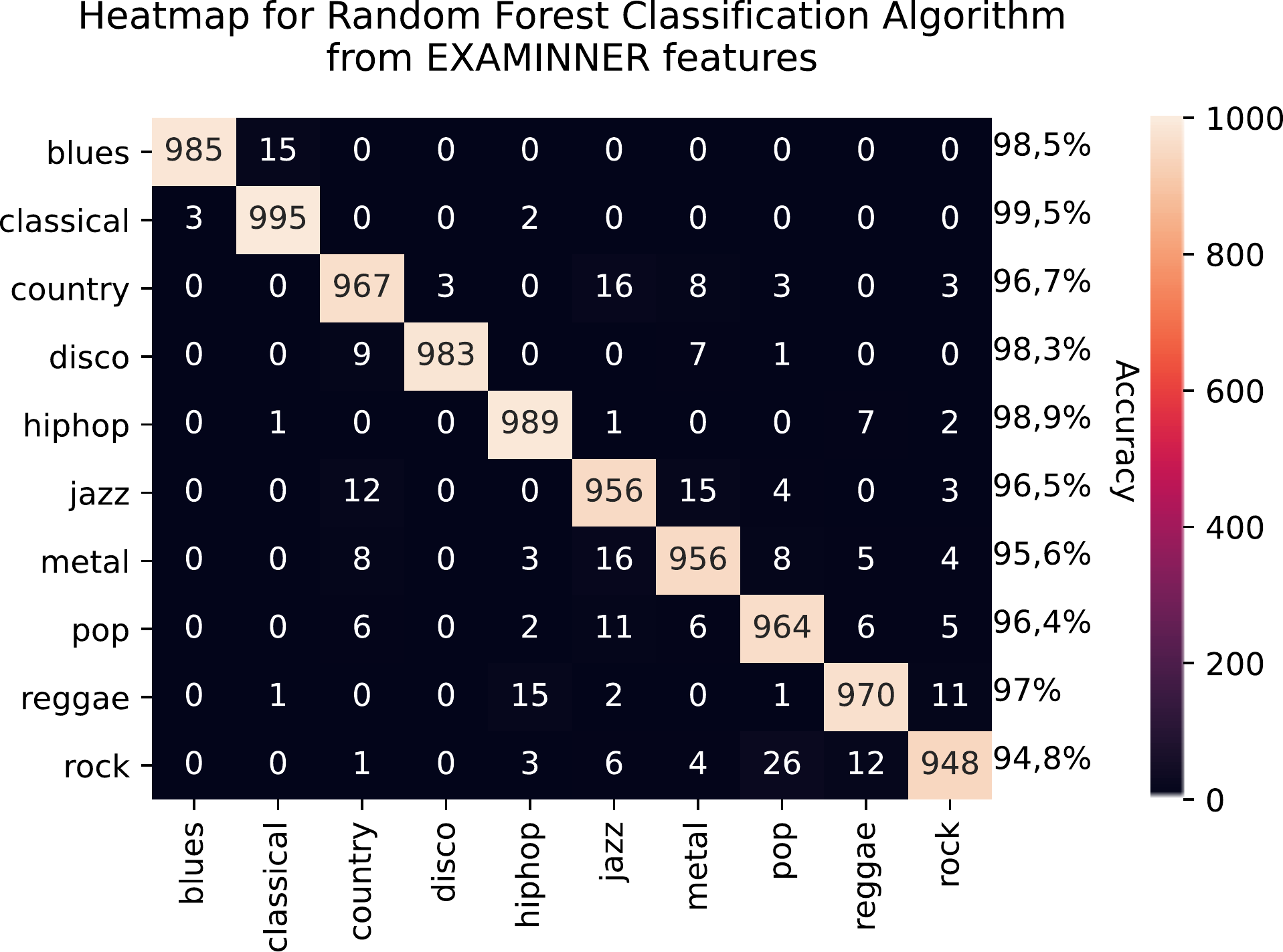}
    \caption{GTZAN dataset (EXAMINNER)}
    \label{fig:gtzanEX}
    \end{subfigure}
    \hfill
    \begin{subfigure}[b]{0.475\textwidth}
    \centering
    \includegraphics[scale=0.3]{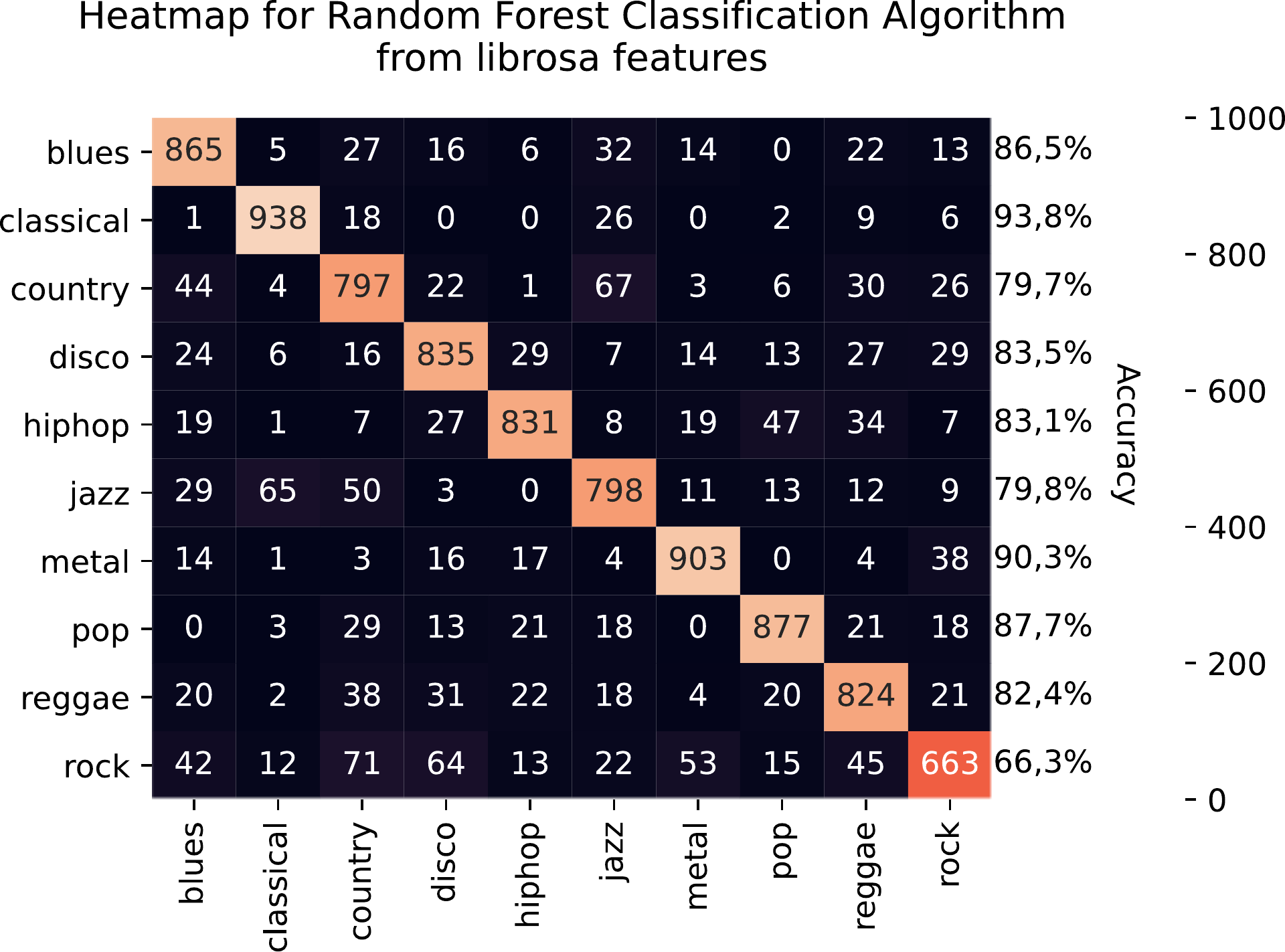}
    \caption{GTZAN dataset (librosa)}
    \label{fig:gtzanlib}
    \end{subfigure}
    \vskip\baselineskip
    \begin{subfigure}[b]{0.475\textwidth}
    \centering
    \includegraphics[scale=0.375]{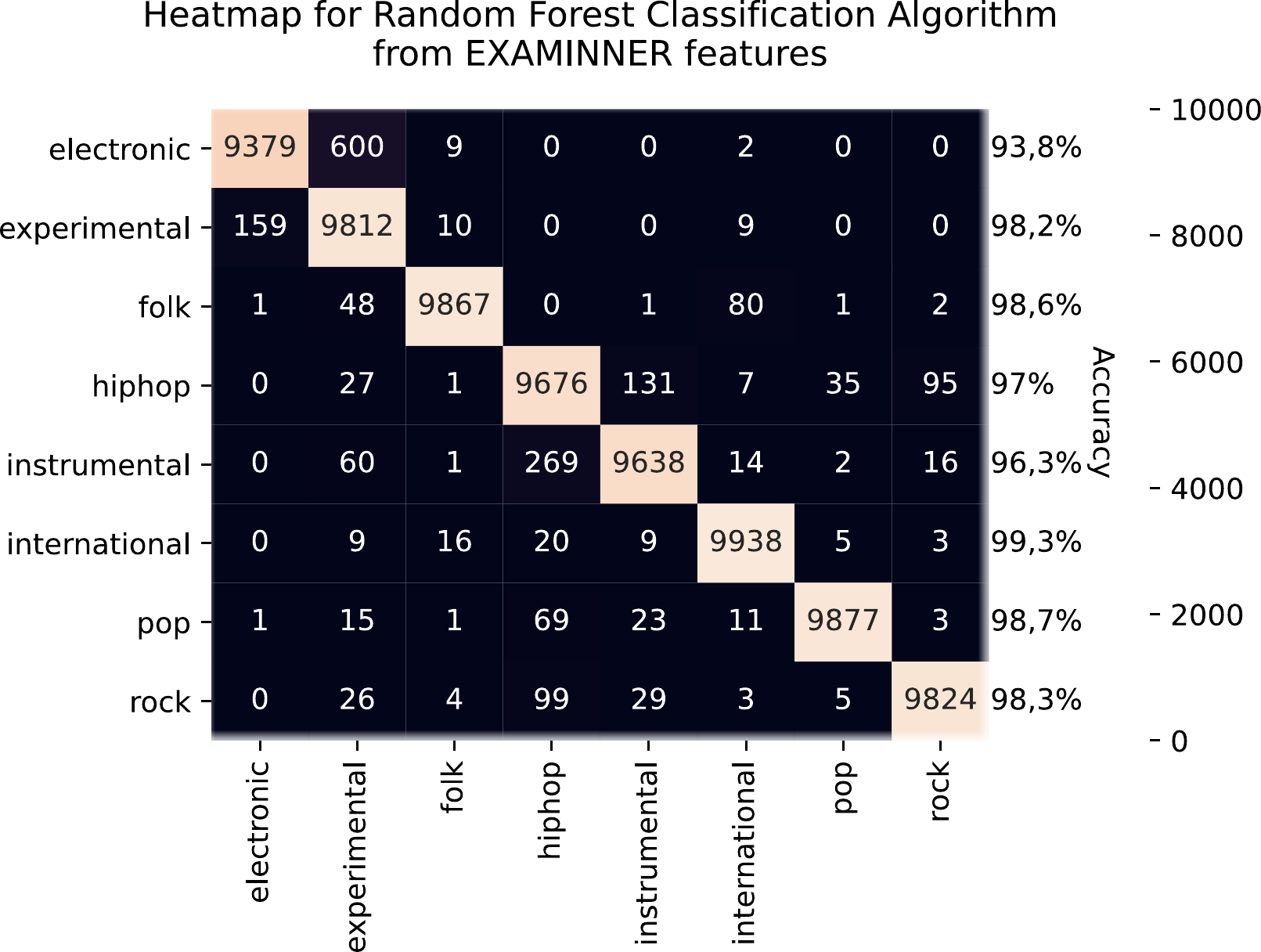}
    \caption{FMA dataset (EXAMINNER)}
    \label{fig:fmaEX}
    \end{subfigure}
    \hfill
    \begin{subfigure}[b]{0.475\textwidth}
    \centering
    \includegraphics[scale=0.375]{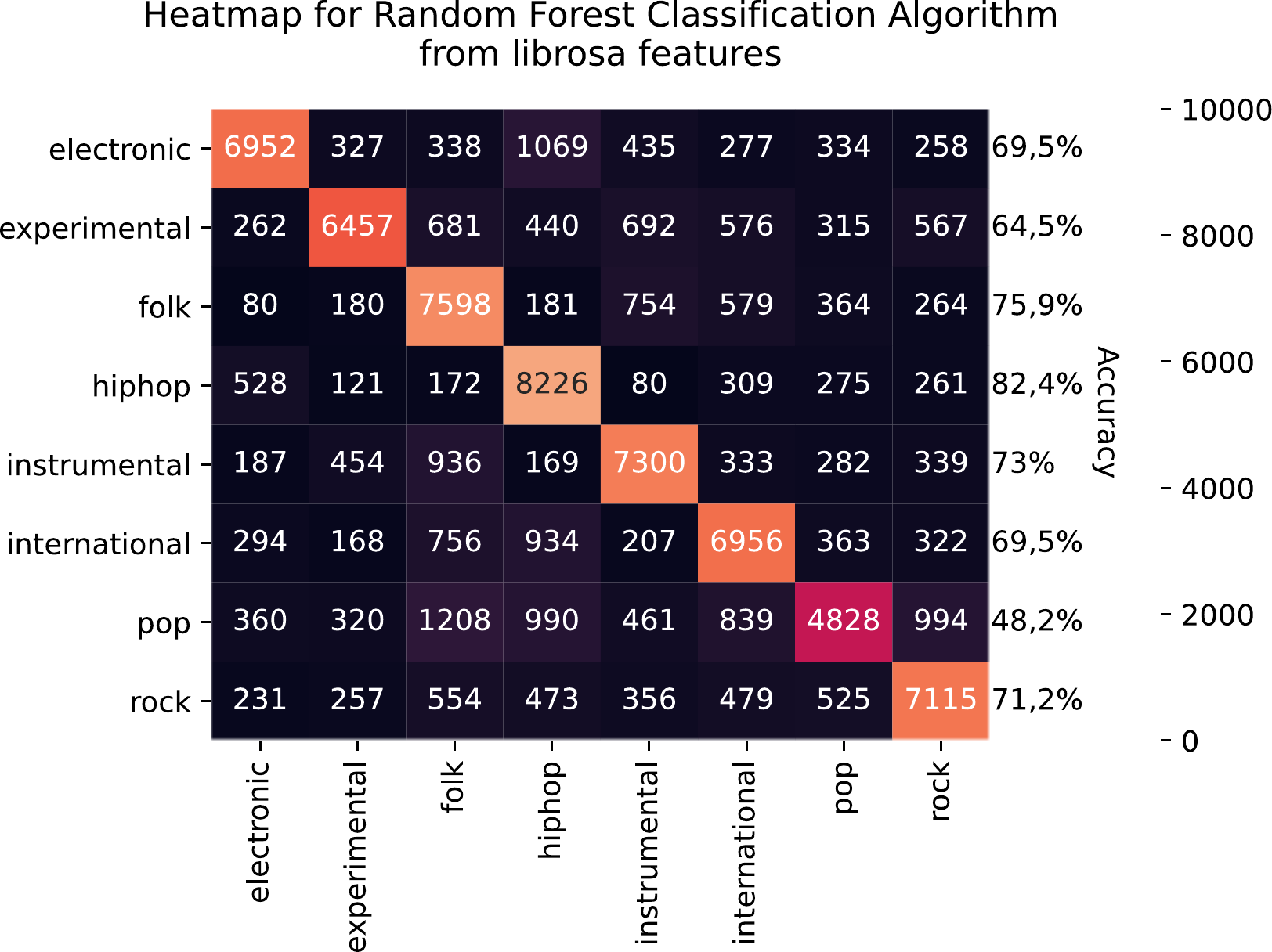}
    \caption{FMA dataset (librosa)}
    \label{fig:fmalib}
    \end{subfigure}
    \caption{Heatmap for the Random Forest classifier considering the features extracted from EXAMINNER and librosa methods.}
    \label{fig:heatmap}
\end{figure*}

Figure \ref{fig:heatmap} shows, for each one of the adopted datasets, the Confusion Matrix of the Random Forest classifier by considering the EXAMINNER and the librosa features. The Random forest classifier was chosen because it was the classifier that presented the best accuracy for the librosa features. Therefore, the aim is to compare the results by considering the best classification of the competitor method to evaluate the EXAMINNER.

It is possible to observe that there is a confusion between electronic and experimental genres, which was noticed in all classification algorithms, showing the proximity of these musical genres and the difficulty in separating them. Regarding the GTZAN dataset, there is a homogeneous classification, i.e. none of the musical genres with a discrepant classification, showing the suitability of the extracted features in the classification process. The relatively slight variation between the errors in the musical genres is expected, given that some genres are composed of similar types of instruments with similar rhythmic patterns \citep{bagci2007automatic}.
The heatmaps for all adopted classifiers produced a similar behavior, which are available at \url{https://github.com/omatheuspimenta/examinner}.

\subsection {Comparing the performance of the proposed approach with literature approaches}

In order to a broaden comparison of the results by the proposed approach with the methods of classification of musical genres available in the literature, a review was performed on the methods that adopted the GTZAN dataset. Table \ref{comparison} presents the comparison of the EXAMINNER feature extraction method and some of the more recent state-of-the-art musical genres classification.
It was observed that several methods applied in GTZAN adopt the SVM classifier. Although EXAMINNER presents more assertive results with other classifiers, as described in the previous section, its results with the SVM classifier were considered, using the linear and polynomial kernels. The aim is to make the results comparable in terms of the features extracted and considered for the same classifier as a basis. This way, the efficiency of the feature extraction method based on the same classifier of the competing methods becomes clearer.

\begin{table*}[htbp]
	\caption{Comparison of the results achieved by the EXAMINNER feature extraction method by considering the SVM classifier with some of the state-of-the-art musical genres classification regarding the GTZAN dataset.}
	\centering
	\begin{tabular}{p{4cm}p{5.5cm}c}
		\hline
		\textbf{Authors} & \textbf{Method} & \textbf{Accuracy} \\ \hline
		
		\cite{Panagakis2} & Locality Preserving Non-Negative Tensor Factorization and Sparse representation-based & 92.4\%  \\ \hline
		
		\cite{Panagakis3} & Bio inspired auditory representation and sparse representation-based & 91.0\% \\ \hline
		
		\cite{Chang} & Multiple feature and compressive sampling based classifier & 92.7\% \\ \hline
		
		\cite{Fu} & Multiple feature combination and support vector machine (SVM) & 90.9\% \\ \hline
		
		\cite{Sigtia} & Deep neural network and Random Forest (RF) with 100 trees & 83.0\% \\ \hline
		
		\cite{nanni2016}  & fusion of acoustic and visual features and support vector machine (SVM) & 83.8\% \\ \hline
		
		\cite{Nanni2017} & fusion of acoustic and visual features and support vector machine (SVM) & 90.7\% \\ \hline
		
		\cite{Choi} & Convolutional neural network and regression (CNN) & 89.8\% \\ \hline
		
		\cite{Bisharad} & Convolutional recurrent neural network (CRNN) & 85.4\% \\ \hline
		 
		EXAMINNER & Complex network measurements and support vector machine (SVM-Linear) & 95.8\%  \\ \hline
		
		EXAMINNER & Complex network measurements and support vector machine (SVM-Poly) & 99.5\%  \\ \hline
		\label{comparison}
	\end{tabular}
\end{table*}

It is possible to highlight that the proposed method achieves the higher average accuracy among competitor methods. The \cite{Panagakis2} present accuracy of $92.4\%$, however the authors have stated that this result was inflated, since this performance is less than that reported \citep{Sturm2,Nanni2017}. In addition, it is important to note that \cite{Chang} and \cite{Fu} reported accuracy of $92.7\%$ and $90.9\%$, respectively. However, their methods are based on manual feature engineering, which are biased by choice of features and might provide higher accuracy \citep{Bisharad}.

Although there are some discussions in the literature about the musical genre classification methods, the proposed method was more accurate than competitor methods regarding the GTZAN dataset. The results can be explained by the fact that the proposed methodology considers the relationships between ``pieces'' of music, represented by musical notes. This relationship of the parts of the music may not be directly identifiable by other feature extraction techniques. The results indicate that the extracted features possibly maps the pattern of adjacency and frequency of musical notes that distinguish the respective musical genres.

Regarding the comparison of results using the FMA dataset, \cite{fma} show that several classifiers are evaluated in order to compare their performances. Using the same features set and six statistical measures, the accuracy of the SVM classifier ranged from 39\% to 63\%, while other classifiers presented lower accuracy. As can be seen in Figure \ref{fig:gtzanresult} (a) and (b), the EXAMINNER presents higher accuracy rates for all applied classification algorithms.

\subsection {Analyzing the features extracted by EXAMINNER}

\begin{figure}[!hb]
	\centering
	\includegraphics[width=0.95\linewidth]{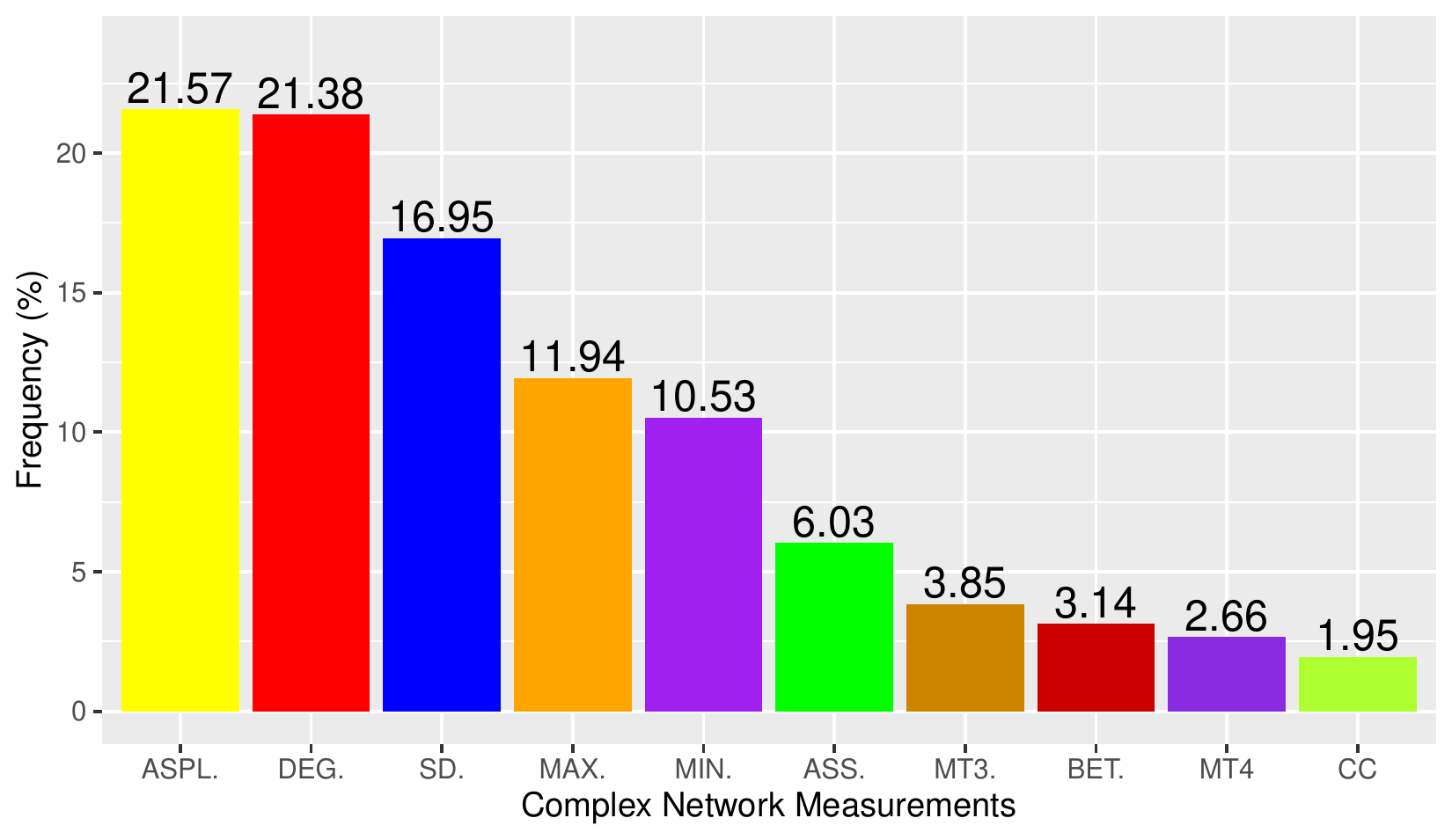}
	\caption{Relative frequency of occurrence of the extracted features in the Random Forest classifier algorithm.}
	\label{frequency}
\end{figure}

In order to analyze which topological measures of complex networks are most relevant, the frequency that each topological measurement was recovered for the musical genre classification. For this analysis, a random forest classifier with 500 trees was adopted, since this number of trees are more robust in relation to 100 trees (default) and the tree structure allows to analyze the features considered in the classification process. Figure \ref{frequency} shows the relative frequency of the use of each measurement (feature) in the musical genres classification process. 

The two topological measures of the complex networks that presented the highest frequency are $ASPL$ and $DEG$, which represent respectively the Average Shortest Path Length and the node degree. The shortest path is the minimum path connecting the node $i$ to the node $j$, that is, the smallest number of edges connecting the node $i$ to the node $j$. An Average Shortest Path Length (ASPL) is defined in (1).
\begin{equation}
	ASPL = \displaystyle \frac{1}{N-1}\sum_{i\neq j}d_{ij},
\end{equation}
\noindent where $d_{ij}$ is the distance between node $i$ and node $j$, and $N$ is the total number of graph nodes. High $ASPL$ values indicate that there is a high distance between musical notes. For example, if the minimum distance between the $A3$ note and the $D2$ note is $7$, it means that there are another $6$ musical notes between the notes; while low values for $ASPL$ represent a sequence of closer musical notes. The Average Shortest Path Length represents this network structure in a general way, indicating the distance between musical notes. The node degree $i$, given by $k_i$, represents the number of edges connected to it. In terms of Adjacency Matrix $(a)$ of an undirected graph, the node degree is given by (2) \citep{costa2007a,BOCCALETTI}.
\begin{equation}
	k_i = \sum_{j}a_{ij} = \sum_{j}a_{ji}
\end{equation}

And the \emph{average degree} is given by (3).
\begin{equation}
	\langle k\rangle = \frac{1}{N}\displaystyle\sum_i k_i = \frac{1}{N}\sum_{ij}a_{ij},
\end{equation}

Musical notes (nodes) with a high number of edges connected to it, represent a hub behavior, being musical notes that concentrate high connectivity in the network. The average standard deviation of the node degrees (SD) is the third most used topological measure in the classification of genres and, according to \citep{costa2007a}, this measure can also be used to characterize networks or phenomena. After that, there are the maximum (MAX) and the minimum (MIN) values, respectively, represented as (4) and (5). These two values also are related to node degree. The 

\begin{equation}
	MAX = \max_i{k_i}
\end{equation}
\begin{equation}
	MIN = \min_i{k_i}
\end{equation}

In addition, the other adopted complex network measurements: assortativity (ASS), frequency of motifs with size 3 (MT3), average betweenness centrality (BET), frequency of motifs with size 4 (MT4) and clustering coefficient (CC) were relevant, however they were selected with lower frequency than the previously described measurements by the classification algorithm. The ASS quantifies the tendency of nodes to connect to nodes that are similar in some way, such as node's degree. The BET quantifies the relevance of a vertex in relation to all the paths of the network, i.e. the more paths pass through a vertex, the greater will be your betweenness. Graph motifs are small connected subgraphs, MT3 and MT4 quantifies the number of motifs with size 3 and 4, respectively in the network. The CC is also known as transitivity, characterize the frequency of loops of order three in the network \cite{BOCCALETTI,costa2007a}.

In this context, the complex networks measurements with the greatest importance are associated with connectivity between network nodes, such as ASPL, DEG, SD, MAX and MIN. More specifically, these measurements are directly associated with the connectivity (degree) and the distance (ASPL) between network nodes (musical notes). On the other hand, the measurements associated with the presence of substructures, such as loops and motifs, or as connection of similar nodes (ASS) and centrality (BET) were selected by the classification algorithm with lower frequency. Thus, the sequence of musical notes and their structural organization in terms of neighborhood were better characterized by topological measurements related to the connection between the musical notes in the network. 

Therefore, the EXAMINNER identifies the adjacent musical notes, counts their frequencies and builds an undirected weighted graph. That is, it maps how many times different adjacent musical notes are repeated and their frequencies. However, instead of mapping one type of repetition at a time, the complex network represents a global map of all frequencies of all adjacent musical notes throughout the music, producing a topological organization of the graph. To summarize and quantify this mapping, measurements of complex networks are extracted from the graph. To identify adjacent patterns at different frequency scales, a threshold is applied. Thus, the method can extract the topological organization generated by musical notes globally and in different scales from the music and these features can distinguish musical genres. In addition, the classification method is able to classify the music in its respective musical genres with high accuracy.

\section{Conclusions}
\label{sec:conclusion}

The classification among musical genres is challenging in face of the extensive amount of new data produced in an unstructured way, which stands out as a matter of interest in Music Information Retrieval. In addition, some musical genres are composed of similar types of instruments with similar rhythmic patterns, increasing the challenge. 

This work presents the EXAMINNER feature extraction method for classification of musical genres based on complex networks and their topological measurements. More specifically, the music is mapped and represented through a complex network. Besides the representation, the method does the characterization by adopting topological measurements of the network. The topological measurements form a feature vector, which is used to classify its musical genre through the analysis of the topological structure formed by the musical notes. In addition, EXAMINNER produces features independently, i.e. not embedded with a classification method, allowing the application of different classification algorithms.

The proposed method was evaluated by adopting the GTZAN dataset, which contains $1,000$ music clips distributed across 10 classes of musical genres and the FMA dataset, which contains $8,000$ musics files distributed across $8$ musical genres. The results of the proposed method were compared with the principal methods available in the literature, using the range of thresholds from 0 to 30 to analyze the results. The accuracy results of the comparison with other methods, when applied to the GTZAN and FMA datasets, showed that the proposed method was more assertive than all other methods. Furthermore, the results show that the representation of music as complex networks can extract features more suitable to its classification than those adopted by the other methods.

Furthermore, the proposed method was suitable for classifying music genres with high accuracy using complex network measurements, which represents a relevant contribution of feature extraction and dimensionality reduction in pattern recognition research, allowing expert music systems to manage and label music content.

As a future work it is suggested an optimization method in order to analyze other parametrization for the word size ($WS$) and the step size ($SS$), that could be more suitable for each type of musical genre, which could lead to a more adequate learning of parameters for each musical genre and as a result, improve the classification.

Finally, the EXAMINNER method was implemented in open source (Python language). The program, the features matrix, figures and the heatmaps (which represent the confusion matrix), as well as all the material necessary for the replication of this work,  are available at \url{https://github.com/omatheuspimenta/examinner}

\section*{Acknowledgements}
This work was funded by Coordena\c{c}\~{a}o de Aperfei\c{c}oamento de Pessoal de N\'{i}vel Superior (CAPES), Conselho Nacional de Desenvolvimento Cient\'{i}fico e Tecnol\'{o}gico (CNPq) (Grant number 406099-2016) and the Funda\c{c}\~{a}o Arauc\'{a}ria e do Governo do Estado do Paran\'{a}/SETI (Grant number 035-2019).


\begin{landscape}
\begin{table}[h!]
\scriptsize
\centering
\begin{tabular}{cccccccccccccccc}
\hline
\textbf{Classifier}   & \textbf{T0} & \textbf{T1} & \textbf{T2}      & \textbf{T3} & \textbf{T4} & \textbf{T5} & \textbf{T6}      & \textbf{T7} & \textbf{T8} & \textbf{T9} & \textbf{T10}     & \textbf{T15}     & \textbf{T20} & \textbf{T25} & \textbf{T30}     \\ \hline
\textbf{Bayes Net}    & 99.50\%     & 99.84\%     & \textbf{99.88\%} & 99.83\%     & 99.85\%     & 99.82\%     & 99.83\%          & 99.79\%     & 99.77\%     & 99.78\%     & 99.71\%          & 99.63\%          & 99.40\%      & 99.24\%      & 99.08\%          \\ \hline
\textbf{MLP}          & 94.78\%     & 95.47\%     & 95.65\%          & 95.70\%     & 95.94\%     & 95.62\%     & 95.23\%          & 95.11\%     & 95.60\%     & 95.04\%     & 95.12\%          & 95.15\%          & 95.84\%      & 95.62\%      & \textbf{96.33\%} \\ \hline
\textbf{Regression}   & 92.96\%     & 92.04\%     & 93.73\%          & 95.68\%     & 96.42\%     & 96.58\%     & 96.62\%          & 96.72\%     & 96.81\%     & 96.80\%     & \textbf{97.01\%} & 96.77\%          & 96.61\%      & 96.68\%      & 96.60\%          \\ \hline
\textbf{RF}           & 94.93\%     & 93.41\%     & 93.59\%          & 95.71\%     & 96.10\%     & 96.18\%     & 96.95\%          & 96.95\%     & 96.92\%     & 97.01\%     & 97.16\%          & \textbf{97.23\%} & 97.10\%      & 97.18\%      & 96.93\%          \\ \hline
\textbf{SVM (Linear)} & 80.49\%     & 86.48\%     & 89.45\%          & 90.43\%     & 91.08\%     & 91.46\%     & 91.71\%          & 92.11\%     & 92.44\%     & 92.50\%     & 92.80\%          & 94.12\%          & 95.05\%      & 95.57\%      & \textbf{95.83\%} \\ \hline
\textbf{SVM (Poly)}   & 91.89\%     & 97.29\%     & 98.76\%          & 99.17\%     & 99.36\%     & 99.46\%     & \textbf{99.54\%} & 99.53\%     & 99.47\%     & 99.47\%     & 99.44\%          & 99.31\%          & 98.96\%      & 98.86\%      & 98.93\%          \\ \hline
\textbf{Average}      & 92.43\%     & 94.09\%     & 95.18\%          & 96.08\%     & 96.46\%     & 96.52\%     & 96.64\%          & 96.70\%     & 96.83\%     & 96.77\%     & 96.87\%          & 97.03\%          & 97.16\%      & 97.19\%      & \textbf{97.28\%} \\ \hline
\textbf{Std Dev}      & 6.40\%      & 4.64\%      & 3.81\%           & 3.35\%      & 3.14\%      & 3.04\%      & 3.00\%           & 2.87\%      & 2.70\%      & 2.74\%      & 2.62\%           & 2.19\%           & 1.72\%       & 1.57\%       & \textbf{1.38\%}  \\ \hline
\end{tabular}
\caption{Average Accuracy of the 10-fold cross validation achieved by the classification algorithms for each adopted threshold considering the GTZAN dataset.}
\label{tab:sup1}
\end{table}


\begin{table}[h!]
\scriptsize
\centering
\begin{tabular}{cccccccccccccccc}
\hline
\textbf{Classifier}   & \textbf{T0}      & \textbf{T1}      & \textbf{T2}      & \textbf{T3} & \textbf{T4} & \textbf{T5}      & \textbf{T6} & \textbf{T7} & \textbf{T8} & \textbf{T9} & \textbf{T10} & \textbf{T15} & \textbf{T20} & \textbf{T25} & \textbf{T30}     \\ \hline
\textbf{Bayes Net}    & 99.40\%          & \textbf{99.44\%} & 99.35\%          & 99.25\%     & 99.19\%     & 99.17\%          & 99.11\%     & 99.05\%     & 98.98\%     & 98.94\%     & 98.88\%      & 98.58\%      & 98.26\%      & 97.86\%      & 97.45\%          \\ \hline
\textbf{MLP}          & 83.08\%          & 83.61\%          & \textbf{83.68\%} & 83.57\%     & 83.33\%     & 82.90\%          & 82.53\%     & 82.61\%     & 82.44\%     & 82.06\%     & 81.95\%      & 80.77\%      & 80.06\%      & 80.13\%      & 80.04\%          \\ \hline
\textbf{Regression}   & 96.05\%          & 95.17\%          & 95.68\%          & 96.30\%     & 96.32\%     & \textbf{96.72\%} & 96.47\%     & 96.21\%     & 95.92\%     & 95.93\%     & 95.91\%      & 95.99\%      & 95.95\%      & 95.81\%      & 95.74\%          \\ \hline
\textbf{RF}           & \textbf{97.58\%} & 95.93\%          & 94.70\%          & 94.94\%     & 94.28\%     & 93.73\%          & 93.97\%     & 93.35\%     & 93.08\%     & 92.69\%     & 92.52\%      & 92.14\%      & 91.12\%      & 90.91\%      & 90.16\%          \\ \hline
\textbf{SVM (Linear)} & 54.11\%          & 59.32\%          & 62.63\%          & 64.41\%     & 65.49\%     & 66.09\%          & 66.44\%     & 66.76\%     & 67.08\%     & 67.26\%     & 67.44\%      & 68.44\%      & 69.20\%      & 70.03\%      & \textbf{70.57\%} \\ \hline
\textbf{SVM (Poly)}   & 66.81\%          & 76.45\%          & 80.51\%          & 82.64\%     & 83.91\%     & 84.74\%          & 85.30\%     & 85.72\%     & 86.15\%     & 86.37\%     & 86.56\%      & 86.92\%      & \textbf{86.99\%}      & 86.93\%      & 86.98\%          \\ \hline
\textbf{Average}      & 82.84\%          & 84.99\%          & 86.09\%          & 86.85\%     & 87.09\%     & 87.22\%          & \textbf{87.30\%}     & 87.28\%     & 87.28\%     & 87.21\%     & 87.21\%      & 87.14\%      & 86.93\%      & 86.94\%      & 86.82\%          \\ \hline
\textbf{Std Dev}      & 18.70\%          & 15.27\%          & 13.63\%          & 12.97\%     & 12.44\%     & 12.22\%          & 12.08\%     & 11.84\%     & 11.65\%     & 11.56\%     & 11.48\%      & 11.18\%      & 10.85\%      & 10.45\%      & \textbf{10.14\%} \\ \hline
\end{tabular}
\caption{Average Accuracy of the 10-fold cross validation achieved by the classification algorithms for each adopted threshold considering the FMA dataset.}
\label{tab:sup2}
\end{table}
\end{landscape}

\bibliography{examinner} 

\end{document}


\begin{center}
\begin{table}[h!]
\centering
\begin{tabular}{|c|c|c|c|c|c|c|c|c|c|c|c|c|c|c|c|c|c|c|}
\hline
\textbf{Classification Method} & \textbf{T0} & \textbf{T1} & \textbf{T2} & \textbf{T3} & \textbf{T4} & \textbf{T5} & \textbf{T6} & \textbf{T7} & \textbf{T8} & \textbf{T9} & \textbf{T10} & \textbf{T15} & \textbf{T20} & \textbf{T25} & \textbf{T30} & \textbf{Best Results} & \textbf{Worst Results} & \textbf{Differences (Best-Worst)} \\ \hline
\textbf{Bayes Net}             & 99,50\%     & 99,84\%     & 99,88\%     & 99,83\%     & 99,85\%     & 99,82\%     & 99,83\%     & 99,79\%     & 99,77\%     & 99,78\%     & 99,71\%      & 99,63\%      & 99,40\%      & 99,24\%      & 99,08\%      & 99,88\%               & 99,08\%                & 0,80\%                            \\ \hline
\textbf{MLP}                   & 94,78\%     & 95,47\%     & 95,65\%     & 95,70\%     & 95,94\%     & 95,62\%     & 95,23\%     & 95,11\%     & 95,60\%     & 95,04\%     & 95,12\%      & 95,15\%      & 95,84\%      & 95,62\%      & 96,33\%      & 96,33\%               & 94,78\%                & 1,54\%                            \\ \hline
\textbf{Regression}            & 92,96\%     & 92,04\%     & 93,73\%     & 95,68\%     & 96,42\%     & 96,58\%     & 96,62\%     & 96,72\%     & 96,81\%     & 96,80\%     & 97,01\%      & 96,77\%      & 96,61\%      & 96,68\%      & 96,60\%      & 97,01\%               & 92,04\%                & 4,97\%                            \\ \hline
\textbf{RF}                    & 94,93\%     & 93,41\%     & 93,59\%     & 95,71\%     & 96,10\%     & 96,18\%     & 96,95\%     & 96,95\%     & 96,92\%     & 97,01\%     & 97,16\%      & 97,23\%      & 97,10\%      & 97,18\%      & 96,93\%      & 97,23\%               & 93,41\%                & 3,81\%                            \\ \hline
\textbf{SVM}                   & 80,49\%     & 86,48\%     & 89,45\%     & 90,43\%     & 91,08\%     & 91,46\%     & 91,71\%     & 92,11\%     & 92,44\%     & 92,50\%     & 92,80\%      & 94,12\%      & 95,05\%      & 95,57\%      & 95,83\%      & 95,83\%               & 80,49\%                & 15,34\%                           \\ \hline
\textbf{Average}               & 92,53\%     & 93,45\%     & 94,46\%     & 95,47\%     & 95,88\%     & 95,93\%     & 96,07\%     & 96,13\%     & 96,31\%     & 96,22\%     & 96,36\%      & 96,58\%      & 96,80\%      & 96,85\%      & 96,95\%      & 96,95\%               & 92,53\%                & 4,42\%                            \\ \hline
\textbf{Standard Deviation}    & 7,15\%      & 4,89\%      & 3,78\%      & 3,34\%      & 3,13\%      & 2,99\%      & 2,95\%      & 2,81\%      & 2,65\%      & 2,69\%      & 2,57\%       & 2,11\%       & 1,65\%       & 1,50\%       & 1,26\%       & 1,26\%                & 7,15\%                 & 5,90\%                            \\ \hline
\end{tabular}
\caption{Average and standard deviation of accuracy to each classifier using GTZAN dataset}
\label{tab:sup1}
\end{table}

\begin{table}[h!]
\centering
\begin{tabular}{|c|c|c|c|c|c|c|c|c|c|c|c|c|c|c|c|c|c|c|}
\hline
\textbf{Classification Method} & \textbf{T0} & \textbf{T1} & \textbf{T2} & \textbf{T3} & \textbf{T4} & \textbf{T5} & \textbf{T6} & \textbf{T7} & \textbf{T8} & \textbf{T9} & \textbf{T10} & \textbf{T15} & \textbf{T20} & \textbf{T25} & \textbf{T30} & \textbf{Best Results} & \textbf{Worst Results} & \textbf{Differences (Best-Worst)} \\ \hline
\textbf{Bayes Net}             & 99,40\%     & 99,44\%     & 99,35\%     & 99,25\%     & 99,19\%     & 99,17\%     & 99,11\%     & 99,05\%     & 98,98\%     & 98,94\%     & 98,88\%      & 98,58\%      & 98,26\%      & 97,86\%      & 97,45\%      & 99,44\%               & 97,45\%                & 1,98\%                            \\ \hline
\textbf{MLP}                   & 83,08\%     & 83,61\%     & 83,68\%     & 83,57\%     & 83,33\%     & 82,90\%     & 82,53\%     & 82,61\%     & 82,44\%     & 82,06\%     & 81,95\%      & 80,77\%      & 80,06\%      & 80,13\%      & 80,04\%      & 83,68\%               & 80,04\%                & 3,64\%                            \\ \hline
\textbf{Regression}            & 96,05\%     & 95,17\%     & 95,68\%     & 96,30\%     & 96,32\%     & 96,72\%     & 96,47\%     & 96,21\%     & 95,92\%     & 95,93\%     & 95,91\%      & 95,99\%      & 95,95\%      & 95,81\%      & 95,74\%      & 96,72\%               & 95,17\%                & 1,55\%                            \\ \hline
\textbf{RF}                    & 97,58\%     & 95,93\%     & 94,70\%     & 94,94\%     & 94,28\%     & 93,73\%     & 93,97\%     & 93,35\%     & 93,08\%     & 92,69\%     & 92,52\%      & 92,14\%      & 91,12\%      & 90,91\%      & 90,16\%      & 97,58\%               & 90,16\%                & 7,42\%                            \\ \hline
\textbf{SVM}                   & 54,11\%     & 59,32\%     & 62,63\%     & 64,41\%     & 65,49\%     & 66,09\%     & 66,44\%     & 66,76\%     & 67,08\%     & 67,26\%     & 67,44\%      & 68,44\%      & 69,20\%      & 70,03\%      & 70,57\%      & 70,57\%               & 54,11\%                & 16,46\%                           \\ \hline
\textbf{Average}               & 86,04\%     & 86,69\%     & 87,21\%     & 87,70\%     & 87,72\%     & 87,72\%     & 87,70\%     & 87,60\%     & 87,50\%     & 87,37\%     & 87,34\%      & 87,18\%      & 86,92\%      & 86,95\%      & 86,79\%      & 87,72\%               & 86,04\%                & 1,68\%                            \\ \hline
\textbf{Standard Deviation}    & 18,97\%     & 16,42\%     & 14,93\%     & 14,31\%     & 13,80\%     & 13,60\%     & 13,46\%     & 13,21\%     & 13,01\%     & 12,92\%     & 12,83\%      & 12,50\%      & 12,14\%      & 11,68\%      & 11,33\%      & 11,33\%               & 18,97\%                & 7,64\%                            \\ \hline
\end{tabular}
\caption{Average and standard deviation of accuracy to each classifier using FMA dataset}
\label{tab:sup1}
\end{table}
\end{center}